\documentclass[aps,a4paper,showpacs]{revtex4}

\usepackage{graphicx}
\usepackage{amsmath,amssymb,color}
\usepackage[english]{babel}

\parskip=\medskipamount

\newcommand{\eq}[1]{(\ref{#1})}
\newcommand{\fig}[1]{Fig.\ref{#1}}

\newcommand{\be}{\begin{equation}}
\newcommand{\ee}{\end{equation}}

\newcommand{\barr}{\begin{array}}
\newcommand{\earr}{\end{array}}

\newcommand{\beqn}{\begin{eqnarray}}
\newcommand{\eeqn}{\end{eqnarray}}

\newcommand{\bs}{\begin{subequations}}
\newcommand{\es}{\end{subequations}}

\newcommand{\bw}{\begin{widetext}}
\newcommand{\ew}{\end{widetext}}

\newcommand{\eps}{\epsilon}

\begin{document}

\title{Fractal folding and medium viscoelasticity contribute jointly to
  chromosome dynamics}





\author{K.E.~Polovnikov $^{1,2}$, M.~Gherardi $^{3,4}$,  M.~Cosentino-Lagomarsino $^{3,5,6}$, M.V.~Tamm$^{1,7}$}

\affiliation{$^1$ Faculty of Physics, Moscow State University,   Moscow, Russia \\
  $^2$ Skolkovo Institute of Science and Technology,  Skolkovo, Russia\\
  $^3$ Universit\`a degli Studi di Milano, Milan, Italy\\
  $^4$ Universit\'e Pierre et Marie Curie, Paris, France\\
  $^5$ CNRS, UMR7238, Paris, France\\
  $^6$ IFOM, FIRC Institute of Molecular Oncology, Milan, Italy  \\
  $^7$ Department of Applied Mathematics, National Research University
  Higher School of Economics,  Moscow, Russia }

\date{\today}

\begin{abstract}
  The chromosome is a key player of cell physiology, and its dynamics
  provides valuable information about its physical organization.
  In both prokaryotes and eukaryotes, the short-time motion of
  chromosomal loci has been described with a Rouse model in a simple or
  viscoelastic medium.  However, comparatively little emphasis has
  been put on the role played by the folded organization of
  chromosomes on the local dynamics.
  Clearly, stress-propagation, and thus dynamics, must be affected by
  such organization, but a theory allowing to extract such information
  from data, e.g. of two-point correlations, is lacking.
  Here, we describe a theoretical framework able to answer this
  general polymer dynamics question, and we provide a scaling
  analysis of the stress-propagation time between two loci at a given
  arclength distance along the chromosomal coordinate. The results
  suggest a precise way to detect folding information from the
  dynamical coupling of chromosome segments.
  Additionally, we realize this framework in a specific theoretical
  model of a polymer with long-range interactions (tuned to make it fold in a fractal way), and
  immersed in a medium characterized by subdiffusive fractional Langevin motion
  (with a tunable scaling exponent), which allows us to derive
  analytical expressions for the correlation functions.
\end{abstract}


\maketitle


The dynamic reorganization of chromosomal DNA plays a fundamental role
in key biological processes at the cellular level, such as
transcription, replication, segregation and
recombination~\cite{Dekker2016,benza2012}.
Measurementes of dynamic fluctuations of chromosomes provide important
evidence on the complex physical nature of the intracellular crowded
medium comprising genome and surrounding medium (bacterial cytoplasm
or eukaryotic
nucleoplasm)~\cite{Bronshtein2016,Tiana2016,Amitai2015,Lagomarsino2015,Kleckner2014}.
Specifically, relevant information comes from tracking chromosomal
loci~\cite{Bronshtein2016,Levi2005,theriot10,theriot12,Bronstein2009,Javer2013,Javer2014}.
Several pieces of evidence indicate that the observed subdiffusion of
tagged loci can be rationalized as a result of the relaxation of Rouse
modes~\cite{Rouse1953,DoiEdwards}, i.e. fluctuations of a Gaussian
chain (see e.g.~\cite{theriot10,Kepten2011}.)
However, the scaling exponents for monomer subdiffusion found
experimentally in different species and conditions may vary, and
generally they differ from the value of 0.5 expected in a simple Rouse
model.

In particular, the medium surrounding the chromosome is reported to
have ``viscoelastic'' properties, intended hereon in the weak sense
that tracer particles show subdiffusive ergodic motion, with an
anti-correlation dip in the velocity-velocity correlation
function~\cite{Lampo2017,theriot10}. The exact physical explanation of
such behavior is unclear, but it might be a consequence of crowding.
The dynamics of a chromosome in such a medium has been
described~\cite{theriot10b,Lampo2016} by coupling Rouse model of
polymer dynamics with forces acting on each monomer following
fractional Langevin equations (see below for more details).
In bacteria, this approach is consistent with the available
experimental evidence on chromosome segregation and subdiffusion of
cytoplasmic particles~\cite{theriot10,theriot12,Lampo2015}.

\begin{figure}[!ht]  \centering
 \includegraphics[width=0.46\textwidth]{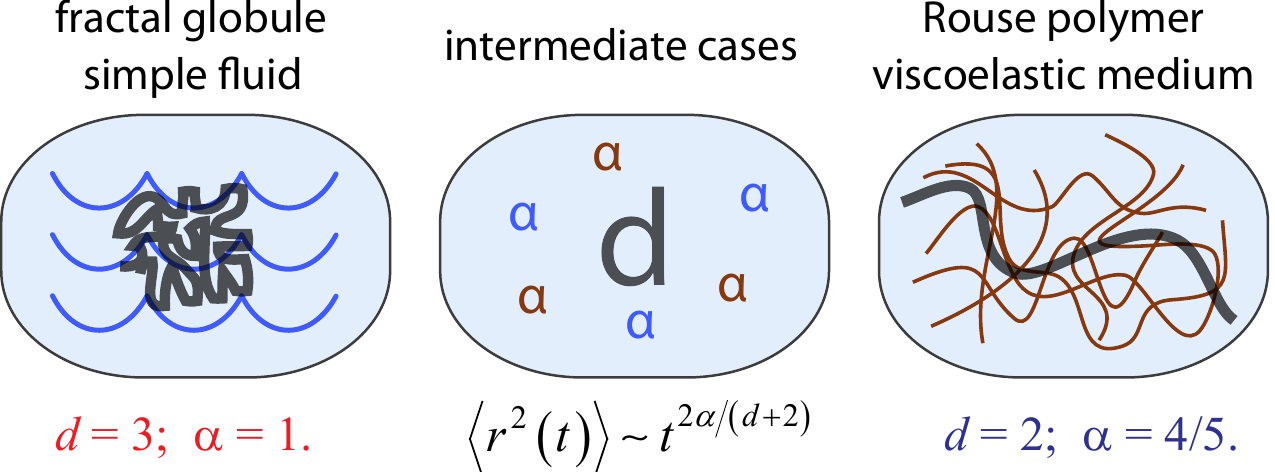}
 \caption{Illustration of the problem. The three panels represent
   different scenarios for a polymer whose folded configuration has a
   generic fractal dimension $d$, immersed in a viscoelastic medium
   characterized by a subdiffusive exponent $\alpha$.  The middle
   panel illustrates the general case studied in this work.  Left
   panel, limiting case of a space-filling fractal ($d=3$) in a simple
   Newtonian fluid. Right panel, limiting case of a Gaussian polymer
   in a viscoelastic medium.  A scaling analysis predicts a
   subdiffusive exponent $2\alpha/(d+2)$ for the segment mean-square
   displacement $\langle r^2(t) \rangle$ in the Rouse regime. Hence, the two limiting
   cases of parameter values illustrated by the left and right panel
   cannot be distinguished by tracking single segments.  }
  \label{fig:1}
\end{figure}

Moreover, the same scenario leads to specific predictions for the case
of two-point correlations between different loci, at arc-length
distance $s$ along the same chromosome~\cite{Lampo2016}, whereby more
information can be extracted.
As the lag time interval increases, the dynamics of loci pairs
changes~\cite{Lampo2016}.
This transition defines a characteristic time scale for the
stress propagation between loci through the polymer backbone.

However, this scenario assumes an oversimplified picture of the folded
state of the chromosome (which influences stress propagation).
Indeed, chromosome packing typically does not follow the Rouse
prediction for the static
exponent~\cite{Imakaev2015,Hofmann2015,Nicodemi2014} and, therefore
(see, e.g.,~\cite{Tamm2015,benza2012}), it is incompatible with the
Rouse dynamics.
Chromosome packing has been described in some cases by a
fractal-globule model~\cite{Grosberg1988,
  Grosberg1993,Lieberman-Aiden2009,Mirny2011,Halverson2014} with
fractal dimension $d = 3$ instead of $d = 2$ for a Rouse chain.
More generally, the chromosome could effectively behave like a fractal
(in a range of scales) with some fractal dimension due to the presence
of a hierarchy of loops caused by bridging and/or loop-extruding
protein complexes, which would affect both its local dynamics and the
propagation of stresses.
%
Figure~\ref{fig:1} illustrates how the stress propagating through the
embedding medium is felt by distant regions along the DNA chain in a
fashion that it is dependent on the dimension of the folded
state. Throughout this Letter we work under the simplifying assumption
that the folded state is (in some range of length scales)
self-similar; thus, its spatial organization can be described by a
single static exponent.

In this Letter, we define a general scaling framework that takes into
account both viscoelastic-like medium generating subdiffusion and
arbitrary dimension of folded structure of a chromosome. We ask about
the relative impact of these two ingredients on the stress-propagation
time between loci pairs at given arclength distance on a chromosome.
Additionally, based on a mathematical model for the Gaussian
self-similar polymer states~\cite{Amitai2013}, we derive analytically
the two-loci correlation functions, which depend on both medium and
packing through their defining parameters. This calculation
confirms our scaling argument and yields precise analytical estimates
for the asymptotic behavior of the system.
The results show how combining single-loci and two-loci tracking
experiments might disentangle the contributions of medium
viscoelasticity from the effects of the folded geometry of a
chromosome.


\paragraph*{Scaling considerations on the joint effects of fractal
  packing and viscoelasticity of the embedding medium.} We consider a
polymer chain whose configuration is described by a function $R(n,t)$
where $n \in [0,N]$ is the (continuous or discrete) coordinate
along the chain, and we assume that the chain conformation is fractal,
i.e.,
\begin{equation}
  \langle R^2 (s,t) \rangle=\langle
  \left( R(n,t)-R(m,t) \right) ^2\rangle
= A s^{2/d}  \ ,
\label{fractal_dimension}
\end{equation}
with some specific fractal dimension $d$. Here $s = |n-m|$ is the
arclength distance. For long chains, and far from the chain ends, the
prefactor $A$ in \eq{fractal_dimension} becomes $m,n$-independent. The
the long-time limit of the Rouse model corresponds to ideal polymer chains with $d=2$, while the
compact fractal globule has $d=3$.
For complex bacterial and eukaryotic chromosomes, the contributions of
(i) incomplete relaxation to equilibrium~\cite{Rosa2008}, (ii) partial
collapse~\cite{Lieberman-Aiden2009,Odi98}, (iii) looped structures due
to bridging proteins and active
enzymes~\cite{Nicodemi2014,Hofmann2015,Scolari2015,Imakaev2015,Dekker2016},
and (iv) branched supercoiled structure due to
plectonemes~\cite{Benedetti2014a} may result in a fractal-like
organization (in a range of length scales) with $d$ between 2 and 3.
The movement of a single locus on a chain can be characterized by the
mean-square displacement
\begin{equation}
  r^2 (t) =\langle \left( R(n,t+t_0)-R(n,t_0) \right)^2 \rangle_{t_0} \sim
  t^{2/z} \ ,
\label{displacement}
\end{equation}
where $z$ is the so-called dynamic exponent and the averaging is over $t_0$.  The standard
argument~\cite{Tamm2015} for a fractal structure in a simple fluid
derives the connection between $z$ and fractal dimension $d$ by
assuming that, due to stress propagation, at lag-time $t$, a region of
spatial size $ x (t) \sim t^{1/z}$ behaves as a single monomer. In the
``free-draining'' limit (negligible hydrodynamic interactions), the
diffusion constant of this coherent region needs to depend on the
number of monomers involved as
$D_{\text{eff}}\sim \left(n(x(t))\right)^{-1} \sim x^{-d}$. Therefore,
the mean-square diplacement of the region is
$D_{\text{eff}} t \sim t^{1-d/z}$.  Since the motion of the whole
region and of a single monomer inside it have to follow the same
behavior, one gets to the consistency condition $2/z = 1 - d /z$,
i.e., $z = 2+d$, the well-known result going back to De
Gennes~\cite{DeGennes1976}. 

Importantly, this generalized Rouse approach relies on the assumption
that chromosome chain dynamics is not restrained by topological
entanglements, which is an open question in the current
literature~\cite{Rosa2008,Bruinsma2014}. A proper description of the
entanglement-dominated regime needs a generalization of the reptation
model~\cite{DoiEdwards} for polymer melts. Note that competing
theories for such a reptation-like dynamics in fractal globules and
ring melts have been suggested recently in the
literature~\cite{Smrek2015,Ge2016}.

\begin{figure}[h]
  \centering
 \includegraphics[width=0.46\textwidth]{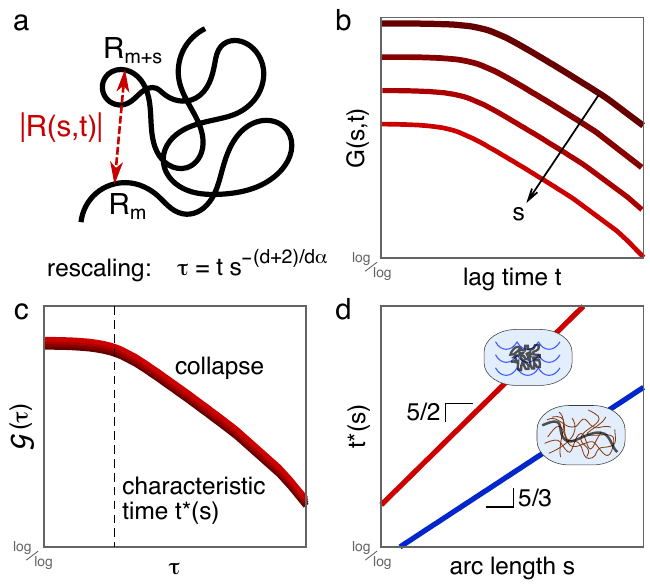}
 \caption{Scaling predictions for monomer-monomer correlations.  A:
   Our calculations quantify stress propagation by the fluctuations of
   the physical distance between monomers at arclength distance
   $s$. B: Sketch of the decay (in log-log scale) of the correlation
   function
   $G(s, t) = \langle \left( R(n,t)-R(m,t) \right)\left( R(n,0)-R(m,0)
   \right) \rangle$,
   for different values of $s$. C: Our scaling analysis predicts the
   collapse of $G(s,t)$ on the master curve ${\cal G}$ upon rescaling
   of time by a characteristic scale
   $t^*(s) \sim s^{\frac{2+d}{\alpha d}}$. For each value of $s$,
   $t^*$ corresponds to a fixed value of $\tau$ in this plot. D: The
   characteristic time $t^*$ is a power law in $s$ (sketched here in
   log-log scale), which distinguishes the two limiting-case scenarios
   illustrated in Fig.~\ref{fig:1}. }
  \label{fig:2}
\end{figure}

Consider now a polymer whose monomers are embedded in a viscoelastic
medium (cytoplasm or nucleoplasm), characterized by a scaling exponent $\alpha$,
so that an isolated tracer particle moves subdiffusively with mean-square
displacement $\langle x^2 (t) \rangle$ growing with time as
$D_{\alpha}t^{\alpha}$ with some $\alpha \leq 1$.
The simplest assumption might be that the effects of polymer
configuration and embedding medium on the monomer mean-square
displacement are factorized, i.e., $R_{n}^2(t)\sim t^{2 \alpha/z}$.
Consider now simultaneous movement of two monomers separated by
arclength distance $s$.  At small times their displacements are
essentially independent, but starting from some typical time, which we
denote $t^*(s)$ their displacements become strongly coupled. This
$t^*(s)$ gives an estimate of the time required for the stress to
propagate between these two monomers. This time can be estimated as a
time at which each monomer moves diffuses a distance comparable to the
spatial distance between the two: $(t^*)^{\alpha/z} \sim
s^{1/d}$. Hence,
\be 
t^* \sim s^{z/(\alpha d)} = s^{(2+d)/(\alpha d)}.  
\label{tstar}
\ee 
The time scale $t^*(s)$ is expected to be measurable from two-point
correlation functions, as described in ref.~\cite{Lampo2016} for
$d=2$.

%

These arguments, however, rely on an arbitrary assumption that the
effects of medium and polymer folding can be factorized.  The
following scaling theory derives the same result based on more general
dimensional grounds (Fig.~\ref{fig:2}).
The monomer displacement and the monomer-to-monomer distance are
expected to be scale invariant until the displacement becomes of order
of the spatial size of the chain. Therefore, the one-locus mean-square
displacement $r^2(t)$ and the two-point correlation function, defined
as
\begin{equation}
\begin{array}{ll}
 G(s, t) =& \medskip \\
  \langle \left( R(n,t+t_0)-R(m,t+t_0) \right)\left(
  R(n,t_0)-R(m,t_0) \right) \rangle_{t_0}&,
 \end{array}
\label{corr}
\end{equation}
are expected to obey the standard scaling forms
\begin{equation}
  r^2 (t) = B t^{2/\zeta}, \; \; G(s,t) = A s^{2/d} {\cal  G} (t s^{-b})
  \label{scaling}
\end{equation}
for the time lags $ t \lesssim N^{\zeta/d} $.
%
%
These conditions imply a scaling relation between $\zeta, d$ and
$b$. Indeed, the scaling hypothesis implies invariance of all
dimensionless quantities under the
transformation $s \to \gamma s, \, t \to \gamma^a s$.
Considering, in particular, $G(s,t)/r^2 (t)$, this
invariance implies $2a/z - 2/d = a - b= 0$, leading to
$\zeta = b d$.

Hence, there is a single independent scaling exponent, which we can
determine by a more rigorous version of the above
argument~\cite{Tamm2015}.
Assume that the physical properties of the medium are described by
a subdiffusive fractional Langevin
equation~\cite{Lampo2016,theriot10b}, so that monomers obey the
equation
\begin{equation}
  \int_0^t dt' K_{\alpha}(t-t') \frac{dR_i(t')}{dt'}
  = f_i (t) + F_i^{\text{polym}}(t),
\label{fractal_langevin}
\end{equation}
where $f_i$ is a random thermal force acting on the $i$-th monomer,
$F_i^{\text{polym}}$ is a force acting on the $i$-th monomer from the
other monomrers of the surrounding chain, and the memory kernel
$K_{\alpha}$ is
\begin{equation}
  K_{\alpha}(t) = D_{\alpha}\frac{(2-\alpha)(1-\alpha)}{|t|^\alpha}
  \label{kernel},
\end{equation}
which reduces to the standard Brownian kernel $K_1(t) = \delta(t)$ for
$\alpha = 1$.
The fluctuation-dissipation theorem implies that the thermal
noise $f(t)$ in the right-hand side of \eq{fractal_langevin} satisfies
$\langle f_i(t)f_i(t') \rangle = K(t-t')$.

Following~\cite{Lampo2016} (and in the spirit of the original Rouse
model), we also assume that the thermal forces acting on different
monomers are uncorrelated, $\langle f_i(t)f_j(t') \rangle =
\delta_{i,j} K(t-t')$.  In this case, the effective diffusion constant
for a group of monomers is inversely proportional to their number, as
assumed above.  Indeed, the equation of motion for the center of mass
$R_\text{c.m.}(n,t) = n^{-1} \sum_{i=1}^n R(i,t)$
of a group of $n$ 
consequential monomers will include a random force of the form $f_{n}
(t) = n^{-1} \sum_{i=1}^n f_i (t)$ whose correlator reads
\begin{equation}
\langle f_{n}(t,s)f_{n}(t') \rangle = \frac{K(t-t')} {n}
=\frac{D_{\alpha}^{\text{eff}}(2-\alpha)(1-\alpha)}{|t-t'|^\alpha}.
\label{d_eff}
\end{equation}
This implies for the monomer displacement
\begin{equation}
  r^2(t) \sim t ^{2/\zeta} \sim D_{\alpha}^{\text{eff}} t^{\alpha}
  \sim  D_{\alpha} t^{ \alpha} t^{- d/\zeta}, 
\end{equation}
thus yielding
\begin{equation}
\zeta = \frac{2+d}{\alpha}; \; b=\frac{2+d}{\alpha d}.
\label{zeta}
\end{equation}
Eq.~\eqref{zeta} is a direct generalization of the results of
ref.~\cite{Lampo2016} for the case of a fractally packed polymer and
of ref.~\cite{Tamm2015} for general viscoelastic embedding medium.
%

\paragraph*{Analytical estimates of the correlation function for the
  beta model in a viscoelastic medium.}
To give more solid grounds to the scaling considerations, we also
defined an explicit model for a specific fractal packing of a polymer,
generalizing the so-called ``beta model''~\cite{Amitai2013}. This is a
Gaussian generalization of the Rouse model that has the advantage of
being mathematically tractable, and we extended it to the case of
viscoelastic embedding medium~\cite{theriot10b}.

The simplest way to define this model is in terms of the behavior of
the Rouse modes $u_p(t),\, p=0...N-1$
\begin{equation}
  u_p(t) =
  \sqrt \frac{1+\delta_{0,p}}{N} \int_0^N ds R(t,s)  \cos
  \frac{p\pi(s-1/2)}{N} \ .
\end{equation}
In the conventional Rouse model, these modes satisfy a set of
independent Ornstein-Uhlenbeck equations
\begin{equation}
  \frac{d u_p}{dt} =
  -\kappa_p u_p + {f}_p; \quad \kappa_p =
  \tau_N^{-1}\sin^2\left( \frac{\pi}{2}\frac{p}{N}\right) \ ,
\label{OU}
\end{equation}
where $\tau_N = a^2/4Dk_bT$ is a relaxation time scale (of order of
the time needed for a monomer  to diffuse by a distance equal to its
own size $a$).

\begin{figure}
 \centering
 \includegraphics[width=0.48\textwidth]{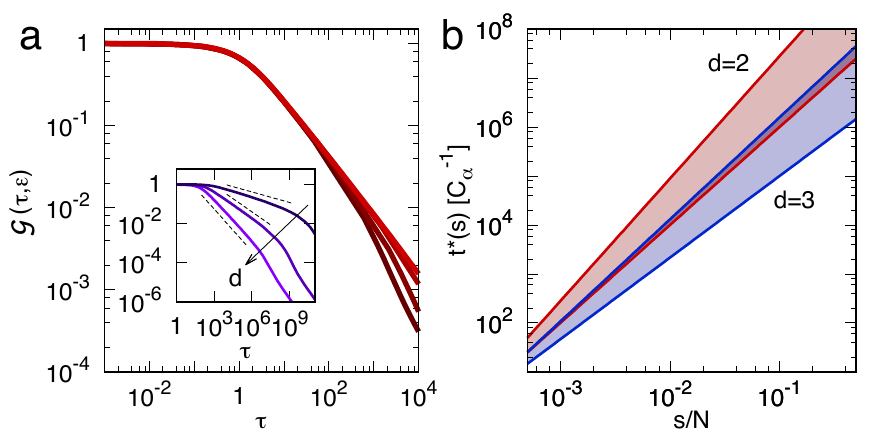}
 \caption{ Analytical predictions from the beta model in viscoelastic
   medium correspond to the scaling expectations.  A: Collapse of the
   two-point correlation function ${\cal G}(\tau,\epsilon)$, given by
   Eq.~\eqref{corr_function_master} as a function of $\tau$.  The red
   curves refer to $d=3$, $\alpha=0.9$ and different values of
   $s$. Inset: change of slope of the master curve ${\cal G}$ with
   varying values of the polymer fractal dimension $d$
   ($d= 4/3 , \ 2, \ 3 \ $, increasing with direction of arrow) at
   fixed $\alpha= 0.9$. The dashed lines are the predictions of
   Eq.~\eqref{intermediate_time}. B: Scaling of $t^*(s)$ (plotted in
   units of $C_\alpha^{-1}$) for $d=2$ (red) and $d=3$ (blue). The
   shaded areas of the two colors correspond to the empirically
   relevant interval, $0.8 < \alpha <1$.  }
  \label{fig:3}
\end{figure}

The beta model is formally defined by replacing the second power of
the sine in $\sin^2\left( \frac{\pi}{2}\frac{p}{N}\right)$ with an
arbitrary power $\beta$. The parameter $\beta$ then plays a role in
the effective description of a fractal folding. In terms of the
original Rouse equation, this corresponds to replacing the second
derivative over the genomic coordinate by a fractional derivative of
order $\beta$ (see Supplementary materials). Introducing a
viscoelastic-like medium in this model corresponds to replacing in
Eq.~\eq{OU} the time derivative of order one with a fractional
derivative as prescribed by Eq.~\eq{fractal_langevin}. As a result,
the Rouse modes now satisfy
\begin{equation}
\int_0^t dt' K_{\alpha}(t-t') \frac{du_p(t')}{dt} = -
\tau_N^{-1}\sin^{\beta}\left ( \frac{\pi}{2}\frac{p}{N}\right) u_p(t) + f_p(t)
\label{fractal_beta1}
\end{equation}
where the memory kernel $K_{\alpha}(t)$ is defined by Eq.~\eq{kernel},
and the random force correlation function is $\langle f_p(t)
f_{p'}(t') \rangle = K(t-t') \delta_{p,p'}$.

%

The connection between the model described by Eq.~\eq{fractal_beta1}
and the scaling considerations above is established by the fact that
regardless of the value of the viscoelastic exponent $\alpha$, the beta-model
converges at long times to an equilibrium state in which
mean-squared monomer-to-monomer distance satisfies
\begin{equation}
\langle (R_n(t)-R_m(t))^2\rangle \sim s^{\beta - 1}, 
\label{monomer1}
\end{equation}
meaning that this equilibrium conformation is fractal, with fractal dimension $d$ 
\be 
d = \frac{2}{\beta - 1}.
\label{betad}
\ee
Additionally, single-monomer displacement scales as
\be
\langle r^2(t) \rangle \sim t^{\alpha(\beta -1)/\beta},
\ee
yielding $ \zeta = \frac{2 \beta}{\alpha (\beta -
  1)} = \frac{2+d} {\alpha}$ which confirms Eq.~\eq{zeta}.

Finally, and most importantly, the solution of Eq.~\eq{fractal_beta1}
allows to compute directly the asymptotics of the scaling function
${\cal G}(\tau)$ defined in Eq.~\eq{scaling} for $N \to \infty$.
Additionally, this calculation can be carried out accounting for
finite-chain-size corrections to the long-chain limit. Indeed (see
Supplementary materials), a more general scaling function
${\cal G}(\tau,\epsilon)$, where $\epsilon = s/N \ll 1$, can be
estimated as
\begin{equation}
  {\cal G}(\tau,\epsilon) \sim
  \sum_{k=1}^{\infty} (-1)^{k+1}
  \frac{\pi^{2k}}{(2k)!} \int_{\epsilon}^{1} x^{2k-\beta} E_{\alpha}
  \left( - \tau^{\alpha} x^{\beta}\right) dx,
\label{corr_function_master}
\end{equation}
where $\tau = C_{\alpha} t s ^{-b} = C_{\alpha} t s^{-\beta/\alpha}$,
$C_{\alpha} = \left(\tau_N \Gamma(3-\alpha) \xi\right)^{-1/\alpha}$ is
an $\alpha$-dependent constant with the dimension of an inverse time,
and $E_{\alpha}(x)$ is the Mittag-Leffler function
\begin{equation}
  E_{\alpha}(x) = \sum_{j=0}^{\infty} \frac{x^j}{\Gamma(1 + \alpha j)},
  \label{ml}
\end{equation}
which reduces to a simple exponential for $\alpha = 1$. Importantly,
the expression in Eq.~\eq{corr_function_master} goes beyond the
approximation of Eq.~\eq{scaling}, as it allows for a small but finite
value of $\epsilon$, while the scaling theory describes only the case
of infinitely long chains ($\epsilon = 0$).

Moreover, it is possible to express the integrals in
Eq.~\eq{corr_function_master} in terms of the 2,2-order Wright
function
\begin{equation}
 _2\psi_2\left[z\bigg|
\begin{matrix}
(a_1, \alpha_1) & (a_2, \alpha_2)  \\
(b_1, \beta_1) & (b_2, \beta_2)
\end{matrix}
\right] = \sum_{k=0}^{\infty}
\frac{ \Gamma(a_1 + \alpha_1 k)\Gamma(a_2 + \alpha_2 k)}{\Gamma(b_1 +
  \beta_1 k) \Gamma(b_2 + \beta_2 k)} \frac{z^k}{k!} \ ,
\label {wright}
\end{equation}
whose asymptotics are known~\cite{Paris2010}; this allows to
distinguish the three following regimes (see Supplementary materials
for technical details):

In the short-time limit $\tau \ll 1$ one gets
\begin{equation}
  {\cal G}(\tau,\epsilon) =  {\cal G}(0,0) \left(1 - A \exp \left(- C \tau^{\alpha} \right)\right)
\label{short_time}
\end{equation}

 At intermediate times,
$\epsilon^{-\beta/\alpha}\gg \tau \gg 1$(this time-scale is much
smaller than relaxation time of the whole chain and the scaling
function in this regime remains $\epsilon$-independent) for the
physically interesting case of $3/2 \leq \beta < 3$ (i.e., $1 < d \leq
4$) the correlations decay as a power-law
\begin{equation}
{\cal G}(\tau,\epsilon)  \sim  \tau^{\alpha(1-3\beta^{-1})}
\label{intermediate_time}
\end{equation}

These two $\epsilon$-independent regimes, visible in \fig{fig:2}, are
in full agreement with the scaling theory. Indeed, Eqs.
\eq{short_time},\eq{intermediate_time} are of the type prescribed by
Eqs.\eq{scaling} and \eq{zeta}, and the crossover at
$\tau^* = t^* s^{-\beta/\alpha} \approx 1$ leads to the estimate
$t^* \sim s^{-\beta/\alpha} = s^{-(d+2)/d\alpha}$ coinsiding with
\eq{tstar}.

Finally, for small but finite $\epsilon$ a third regime arises for
$\tau \gg \epsilon^{-\beta/\alpha}$. 
It corresponds to the relaxation of the whole chain, and is akin to the
behavior of single particle correlation functions in simple (for
$\alpha = 1$) and viscoelastic (for $\alpha<1$) medium:
\begin{equation}
\begin{array}{rlll}
{\cal G}(\tau,\epsilon) &\sim &\epsilon^{1 - 2\beta } \exp \left(- C \tau \epsilon^{\beta}\right)\; &\text{ for }\alpha = 1 \medskip \\
{\cal G}(\tau,\epsilon) &\sim & \epsilon^{1-2\beta} \tau^{-\alpha}\; &\text{ for }\alpha < 1
\end{array}
\label{long_time}
\end{equation}





In conclusion, the framework combining the beta model with the
approach of Spakowitz and coworkers~\cite{theriot10b,Lampo2016}
provides a valuable building block for the description of the motion
of both eukaryotes chromatin and bacterial
chromosomes~\cite{Zidovska2013,Bruinsma2014,Kleckner2014,Pederson2015},
which can be generalized to the case of nonequilibrium
fluctuations~\cite{Vandebroek2015}.

The general picture defined here depicts a wider scenario for the
stress propagation than the one recently proposed by Lampo et
al.~\cite{Lampo2016}.  The most important prediction is that a joint
measurement of two functions $r^2 (t)$ and $G(s, t)$ allows to
disentangle the effects of chain organization and embedding medium.
Indeed, two independent measurements of the exponents $z$ (from
single-loci MSD) and $b$ (from two-loci correlations) allows to
reconstruct $\alpha$ and $d$ as follows
\begin{equation}
  d = \frac{1} {z b}; \; \; \alpha = \frac{2 z b+1}{b} \ .
\end{equation}

Finally, we note that the results obtained here, as well as those of
ref.~\cite{Lampo2016}, are concerned with a polymer embedded in a
complex medium whose reaction is non-local in time but local in space,
i.e.  forces acting on different monomer units are assumed to be
uncorrelated. It seems natural to expect that this locality in space
may in fact break down in some viscoelastic media, and generalization
of the results obtained above for such a spatially non-local case is
an important and challenging task.

\section*{Acknowledgments}
This work was partly financially supported by RFBR grant 14-03-00825,
and EU-Horizon2020 IRSES project DIONICOS (612707). MCL and MG
acknowledge support from the International Human Frontier Science
Program Organization, grant RGY0070/2014. Authors are grateful to
R.~Metzler and S.~Nechaev for many interesting discussions, to
A.~Grosberg for valuable critical comments, 
and to M.~Baiesi, C.~Vanderzande and Y.~Garini for feedback on this
manuscript.

\section*{Supplementary materials}
\subsection{Rouse model}

The standard Rouse model takes into account only interactions with the
neighbours along the polymer chain and does not consider any excluded
volume effects~\cite{DoiEdwards, GrosbergKhokhlov, Rubinstein}. In
particular, it ignores any hydrodynamic interations between monomers.
In the Rouse description the interaction of polymer chain with the
solvent is strictly local, which is expressed mathematically by the
fact that the random noise representing the thermal force acting on
monomers is delta-correlated in space and time.

Under these simplifying assumptions, the Langevin equation of the system reads
\be
\xi \frac{d R_n}{dt} = - \frac{\partial}{\partial R_n} \phi(R_1, ..., R_N) + F_n
\label{langevin_1}
\ee
where $\xi$ is a friction coefficient, the potential energy of monomer-monomer interactions $\phi(R_1, ..., R_N)$ is
\be
\phi(R_1, ..., R_N) = \frac{k}{2}\sum_{i = 1}^{N-1} (R_i - R_{i+1})^2; \quad k = \frac{Dk_BT}{a^2}
\label{potential}
\ee
and the random thermal force  $F_n$ satisfies
\be
\langle F_n(t) F_m(t') \rangle = 2D k_B T \xi \delta(t - t') \delta_{n,m}
\ee
here $D$ is the space dimension.


Substituting \eq{potential} into \eq{langevin_1} one gets Langevin equation in the form
\be
\xi \frac{d R_n}{dt} = k(R_{n+1} + R_{n-1} - 2R_n) + F_n,
\label{Rouse}
\ee
which corresponds in the continuous limit to the Edwards-Wilkinson equation of the form
\be
\xi \frac{\partial R(n, t)}{\partial t} = k\frac{\partial^2 R(n, t)}{\partial n^2} + F(n, t)
\ee

Equation \eq{Rouse} can be diagnoalized by Fourier transform $\{R_n\}_1^N \to \{u_p\}_0^{N-1}$:
\be
\alpha_p^{(n)} =
    \begin{cases}
        \sqrt \frac{2}{N} \cos \frac{p\pi(n-1/2)}{N}, p = 1, 2, .., N-1 \\
        \sqrt\frac{1}{N}, p = 0
    \end{cases}
\ee
\be
R_n(t) = \sum_{p=0}^{N-1} u_p(t) \alpha_p^{(n)}; \quad u_p(t) = \sum_{n=1}^{N} R_n(t) \alpha_p^{(n)}
\label{Fourier}
\ee

In these new coordinates the Langevin equations decouple, and the resulting equations are of Ornstein-Uhlenbeck type:

\be
\xi \frac{d u_p}{dt} = -\kappa_p u_p + {f}_p; \quad \kappa_p = 4k\sin^2 \frac{p\pi}{2N} = \frac{4Dk_BT}{a^2}\sin^2\left( \frac{\pi}{2}\frac{p}{N}\right)
\label{OU}
\ee
the corresponding relaxation time $t_p$ of a  $p$-th Rouse modes is
\be
t_p = \frac {\xi} {\kappa_p}; \; \;
t_1 = t_R = N^2 \frac{\xi a^2}{\pi^2 D k_BT} = N^2 t_N
\ee
where the longest relaxation time $t_1$ is also often called the Rouse time of the chain $t_R$, and the microscopic timescale is defined by $t_N$ - a typical time it takes for a single monomer unit to diffuse at a distance equal to the typical distance between monomers $a$.

The random force in \eq{OU} is
\be
{f}_p(t) =  \sum_{n=1}^{N} F_n(t) \alpha_p^{(n)}; \;\; \langle {f}_p(t){f}_q(t') \rangle = 2D k_B T \xi \delta(t - t') \delta_{p,q}
\ee

The Rouse potential in normal coordinates adopts the form
\be
\phi(u_0, ..., u_{N-1}) = \frac{1}{2}\sum_{p=0}^{N-1} \kappa_p u_p^2
\label{potential_normal_1}
\ee

For $p=0$ \eq{OU} gives the equation on the center-of-mass displacement, which, since $\kappa_0 = 0$, reduces to the overdamped equation for Brownian motion with the friction coefficient $\xi_0 = \xi N$ and initial condition $u_0(0) = 0$.

For $p>0$ the formal solution of \eq{OU} reads
\be
u_p(t) = u_p(0)\exp(-t/t_p) + \xi^{-1}\int_0^t f_p(t') \exp \left(-\frac{t-t'}{t_p}\right) dt',
\ee
at times much larger than $t_p$ the solution becomes independent of initial conditions and the mean-square displacement coverges to
\be
\lim_{t \to \infty} \langle u_p^2(t) \rangle = \xi^{-2}\int_{-\infty}^t \int_{-\infty}^t \langle f_p(t') f_p(t'')\rangle \exp\left(-\frac{t-t'}{t_p}\right) \exp\left(-\frac{t-t''}{t_p}\right) dt' dt'' =  \frac{D k_B T}{\kappa_p},
\label{normal_long-time}
\ee
on the other hand, for $t \ll t_p$ the mean-square displacement of $u_p$ grows linearly with time as for normal diffusion:
\be
\langle (u_p(t) - u_p(0))^2 \rangle_{t \ll \tau_p} \sim \xi^{-2}\int_0^t \int_0^t\langle f_p(t') f_p(t'')\rangle dt' dt'' =  \frac {2D k_B T} {\xi} t
\label{normal_short-time}
\ee

\subsection{Beta model}

In the $t \gg t_R$ regime, the conformational statistics of a Rouse
polymer chain converges to that of an ideal Gaussian polymer chain. In
order to describe the dynamics of a chain with different equilibrium
statistics (e.g., a swollen coil or a fractal globule) in the similar
framework one needs to modify the Rouse equations in order to include
the interactions between the monomers.

One possible way to do that is the so-called beta model introduced in
ref.~\cite{Amitai2013} which, despite having no clear microscopic
justification, has the advantage of describing parametrically the
fractal structure of the polymer conformations and at the same time
being analytically tractable. The basic idea of the beta-model is to
modify the structural potential $\phi(R_1, ..., R_N)$ by introducing
effective springs between monomers positioned arbitrarily far from
each other along the chain. Monomer-monomer interactions in this model
are relatively weak, but infinitely-ranged (contrary to typical
molecular systems, where these interactions are strong but
short-ranged). However, the beta model captures phenomenologically the
main large-scale properties of a fractally organized polymer in the
mean-field sense (this reasoning can be compared to arguments in
support of using soft potentials in DPD simulations~\cite{Groot1997}).

If all the monomers are connected to each other by a set of springs,
the generalized Rouse potential of the chain should read:
\be
\tilde{\phi}(R_1, ..., R_N) = \frac{1}{2} \sum_{i>j}a_{ij} (R_i - R_j)^2 = \frac{1}{2}\sum_{i=1}^{i=N} R_i^2 \sum_{j \ne i} a_{ij} - \sum_{i>j} a_{ij} R_i R_j,
\label{beta_potential}
\ee
the corresponding Langevin equation being
\be
\xi \frac{d R_n}{dt} = \sum_{m \ne n}a_{mn} (R_m - R_n) + F_n,
\label{Rouse}
\ee

Beta model \cite{Amitai2013} corresponds to a particular choice of $a_{ij}$ such that in the same normal coordinates $u_p$ \eq{Fourier} the generalized potential takes the form similar to \eq{potential_normal_1}:
\be
\tilde{\phi}(u_0, ..., u_{N-1}) = \frac{1}{2}\sum_{p=0}^{N-1} \tilde {\kappa}_p u_p^2
\ee
but with modified $\beta$-dependent eigenvalues
\be
\tilde {\kappa}_p = \frac{4 D k_BT}{a^2}\sin^{\beta}\left( \frac{\pi}{2}\frac{p}{N}\right)
\ee
(thus $\beta = 2$ corresponds to the regular Rouse model).

Using the inverse Fourier transform, one gets the following results for the original parameters $a_{ij}$ of the potential:

\be
\tilde{\phi}(R_1, ..., R_N) = \frac{1}{2} \sum_{p=0}^{N-1} \tilde{\kappa}_p \left(\sum_{n=1}^{N} R_n \alpha_p^{(n)} \right)^2 =
\frac{1}{2} \sum_{l,m = 1}^N A_{lm} R_lR_m,
\label{newA}
\ee
where
\be
A_{lm} = \sum_{p=0}^{N-1} \tilde{\kappa}_p \alpha_p^{(l)} \alpha_p^{(m)} = 4k \frac{2}{N} \sum_{p=1}^{N-1} \sin^\beta \left(\frac{p\pi}{2N}\right)
\cos\left(\left(l-\frac{1}{2}\right)\frac{p\pi}{N}\right)\cos\left(\left(m-\frac{1}{2}\right)\frac{p\pi}{N}\right)
\label{inverse}
\ee

Now comparing \eq{beta_potential} and \eq{newA} one can notice that $A_{ll} = \sum_{j \ne l} a_{lj}$ and $A_{lm} = -2 a_{lm}$ for $l \ne m$.
If we insert $\beta = 2$ into \eq{inverse}, it readily gives $A_{lm} = 2(\delta_{l-m, 0} - \delta_{|l-m|, 1})$ getting us back to the Rouse model.

Similarly to the Rouse model, in the normal coordiantes the relaxation of the beta-model can be understood in terms of a set of simultaneously relaxing Ornstein-Uhlenbeck oscilators with relaxation time corresponding to the $p$-th mode being equal to $ \xi \tilde{\kappa}_p^{-1}$. For long-range modes with $p\ll N$ the relaxation times can be approximated by $ \frac{\xi a^2}{k_B T} \left(\frac{N}{p}\right)^{\beta}$, while in the traditional Rouse model the corresponding times are$ \frac{\xi a^2}{k_B T}  \left(\frac{N}{p}\right)^2$. 

\subsection{Beta model in fractal environment and single monomer displacement}

Consider now the behavior of the beta-model in a media where the
monomers are subject to the generalized Langevin equation defined by
equation (6) of the main text with $F_{\text{polym}}$ being
the derivative of the beta-model potential, \eq{newA}. Changing
variables to the normal coordinates once again allows to decouple of
the equations leading to: 
\be 
\xi_{\alpha} \int_0^t dt' K(t-t')
\frac{du_p(t')}{dt} = -\tilde {\kappa}_p u_p(t) + \tilde{F}_p(t) 
\ee
where eigenvalues
$\tilde {\kappa}_p = 4k\sin^\beta \left(\dfrac{p\pi}{2N}\right)$ 
and
\be \langle \tilde{F}_p(t)\tilde{F}_q(t') \rangle = \xi_{\alpha} k_B T K(t-t')
\delta_{p,q} \ee

The solution of this generalized Ornstein-Uhlenback problem can be
obtained as follows.  The correlation function of a $p$-th normal
coordinate $C_p(t) = \langle u_p(t) u_p(0) \rangle$, satisfies:

\be \xi_{\alpha} \int_0^t dt' K(t-t') \frac{dC_p(t')}{dt} = -\tilde {\kappa}_p
C_p(t), \ee where we use the fact that the average noise is null,
$\langle \tilde{F}_p(t) \rangle = 0$.  The solution of this equation
is known to be~\cite{Metzler2000}
\be C_p(t) = C_p(0)
E_{\alpha}\left[-\left(\frac{t}{\tilde {t}_p}\right)^\alpha
\right]
\label{ml}
\ee
where $E_{\alpha}(x)$ is the Mittag-Leffler function \cite{Samko}: 
\be
E_{\alpha}(x) = \sum_{j=0}^{\infty} \frac{x^j}{\Gamma(1 + \alpha j)},
\label{mitlef}
\ee and 
\be
\tilde {t}_p = \left(\frac{\Gamma(3-\alpha) \xi}{\tilde {\kappa}_p}\right)^{1/\alpha}.
\ee
Note
that the rightful dimensionality of time is achieved by a proper choice of 
dimensionality for generalized friction coefficient, $\xi_{\alpha}$, in the generalized 
Langevin equation.

At times much larger than $\tilde {t}_p$, the average values of the
normal coordinates $\langle u_p(t) \rangle$ converge to zero, while
their average squares $\langle u_p^2(t) \rangle$ converge to their
equilibrium values, $D k_B T\tilde{\kappa}_p^{-1}$, dictated by the
equipartition theorem.

Consider now the monomer displacement $\langle r_n^2 (t) \rangle$ for some given $n$ as a function of time which is a weighted sum of $C_p(t)$ with fixed coefficients of order 1. Assume also that we start observation at the equilibrated state so that average initial values of the normal coordinates are the same as their long-time limits:
\be
\overline{ u_p^2(0) } = \overline{C_p(0)} = \lim_{t\to\infty} \langle u_p^2(t) \rangle = D k_B T\tilde{\kappa}_p^{-1},
\label{CP}
\ee
where bar designates averaging over initial conditions, while angular brackets designate, as usual, averaging over the realization of the noise. Now, if the sum of $\overline{C_p(t)}$ is controlled by the terms with $1 \ll p \ll N$ (i.e.. for times $\tilde{t}_N \ll t \ll \tilde {T}_1$ one can replace it with the integral
\be
\langle r_n^2 (t) \rangle = \sum_{p=1}^N 2(\overline{C_p(0)} - \overline{C_p(t)}) \left(\alpha_p^{(n)}\right)^2 \sim \int_1^N (\overline{C_p(0)} - \overline{C_p(t)}) dp = \int_1^N D k_B T\tilde{\kappa}_p^{-1} \left(1- E_{\alpha}\left[-\left(\frac{t}{\tilde {t}_p}\right)^\alpha \right] \right) dp
 \ee
 and after changing variables to
 \be
x = \left(\frac{t}{\tilde{t}_p}\right)^{\alpha};\;\;\;
\tilde{\kappa}_p = 4k (\pi p/ 2 N)^{\beta} = x \frac{\xi\Gamma(3-\alpha)}{t^\alpha};\;\;\; p = \frac{2 N}{\pi} \left(\frac{x \xi\Gamma(3-\alpha)}{4k}\right)^{1/\beta} t^{\alpha/\beta}
\ee
one gets
\be
\langle r_n^2 (t) \rangle =  \frac{k_B T}{\xi} N \left(\frac{\xi}{k}\right)^{1/\beta} t^{\alpha - \alpha/\beta} \times \text{(some dimensionless integral)}
\ee
Interestingly, this equation implies that the exponent governing the
time dependence of $\langle r_n^2 (t) \rangle$ is just a product of
exponents for a free particle in viscoelastic fluid and the exponent
for the beta-model in the standard Brownian regime (in particular, for
$\beta = 2$ one recoveres the result for the Rouse model in
viscoelastic environment \cite{Lampo2016}).

\subsection {Monomer-monomer correlations in the beta model}

Let us now consider the correlations of two monomer positions in the
equilibrium state of the beta-model. The beta model (as well as the
Rouse model) implies that the long-time limit solution converges to
some distinct equilibrium state. For the Rouse model this state is
known to be the ideal chain state (see
\cite{Tamm2015}). We now discuss the equilibrium state associated to
the beta model.

We analyze the autocorrelation function of monomer-to-monomer distance $R_{nm}(t) = R_n(t) - R_m(t)$:
\be
\langle R_{nm}(t)R_{nm}(0) \rangle = \sum_{p=0}^{N-1} \langle u_p(t)u_p(0) \rangle (\alpha_p^{(n)}-\alpha_p^{(m)})^2
\ee
In the equilibritum state the products of Rouse coordinates are averaged over both the initial state and the realization of disorder, so combining \eq{ml},\eq{mitlef},\eq{CP} one gets
\be
\langle u_p(t)u_p(0) \rangle = \overline{C_p(t)}  = \overline{C_p(0)}E_{\alpha}\left[-\left(\frac{t}{\tilde{t}_p}\right)^{\alpha}\right]
\ee


Substituting the Fourier coefficients, one obtains:
\be
\langle R_{nm}(t)R_{nm}(0) \rangle \sim a^2 N^{-1} \sum_{p=1}^{N-1} \sin^{-\beta} \left( \frac{\pi p}{2 N}\right) \sin^2 \left( \frac{\pi p |n-m|}{2 N}\right) \sin^2 \left( \frac{\pi p (n+m)}{2 N}\right) \overline{C_p(t)}
\label{triple_sine}
\ee
Except for the monomers very close to the end of the chain, i.e. $(n+m) \ll N$ or $(2N-n-m)\ll N$), the factor $\sin^2 \left( \dfrac{\pi p (n+m)}{2 N}\right)$ is a rapidly oscilating function of $p$ and, when replacing summation with integration we will simply approximate it by $1/2$. The main contribution into the remaining sum
\be
\langle R_{nm}(t)R_{nm}(0) \rangle \sim a^2 N^{-1} \sum_{p=1}^{N-1} \sin^{-\beta} \left( \frac{\pi p}{2 N}\right) \sin^2 \left( \frac{\pi p |n-m|}{2 N}\right) \overline{C_p(t)}
\ee
comes from the first half-period of the second multiplier, $1 \leq p \leq \frac{N}{|n-m|}$.
Within this interval $\frac{p}{N} \le \frac{1}{|n-m|}\ll 1$ and the first sine can be replaced by $\left( \pi p/2 N\right)^{-\beta}$. Switching from summation to integration over variable $\lambda=p/N$ and expanding the second sine to the Taylor series, we have:
\be
\langle R_{nm}(t)R_{nm}(0) \rangle = G(s,t) \sim \frac{a^2}{2}\left(\frac{2}{\pi}\right)^{\beta}\sum_{k=1}^{\infty} (-1)^{k+1}\frac{\pi^{2k}}{(2k)!}  s^{2k}  I_{\alpha}(k, s,t)
\label{result}
\ee
where, similarly to the main text we introduced $s=|n-m|$ and the notation $G(s,t)$ for the correlation function. The integral $I_{\alpha}(k, s, t)$ is defined as
\be
I_{\alpha}(k, s, t) = \int_{N^{-1}}^{s^{-1}} \lambda^{2k-\beta} E_{\alpha}\left(-\lambda^{\beta}
\left(\frac{t}{\tilde{t}_N}\right)^\alpha \right) d\lambda
\label{f}
\ee
for $\alpha < 1$ (fractal medium), and
\be
I_{1}(k, s, t) = \int_{N^{-1}}^{s^{-1}} \lambda^{2k-\beta} \exp(-t\lambda^{\beta}/\tilde{t}_N) d\lambda
\label{nf}
\ee
for $\alpha=1$ (Newtonian medium).
Note, that since $k \ge 1$ these integrals \eq{nf}-\eq{f} converge for all $\beta < 3$.

In order to study the behavior of the correlation function at short times we expand the Mittag-Leffler function in \eq{f} into Fourier series\cite{Samko}. Changing the order of summation and integration, leads to the following series representation of the integrals in the limit $N \to \infty$:
\be
I_{\alpha}(k, s, t) = s^{\beta-1-2k} \sum_{n=0}^{\infty} \frac{1}{\Gamma(\alpha n + 1) (2k + \beta(n-1) + 1)} \left(-\frac{t^{\alpha}}{\tilde{t}_N^\alpha s^{\beta}}\right)^n
\label{f1}
\ee
Substituting this representation into autocorrelation function \eq{result} leads to:
\be
G(s,t) \sim a^2 s^{\beta-1}\sum_{n=0}^{\infty}
\frac{1}{\Gamma(\alpha n + 1)}\left(-\frac{t^{\alpha}}{\tilde{t}_N^\alpha s^{\beta}}\right)^n \sum_{k=1}^{\infty}(-1)^{k+1}\frac{\pi^{2k}}{(2k)!} \frac{1}{2k + \beta(n-1) + 1}
\label{res1}
\ee
One sees therefore that at $t=0$
\be
G(s, t=0) = \langle R_{nm}^2(0)\rangle \sim a^2 s^{\beta-1}
\ee
independently of $\alpha$, i.e., the equilibrium state corresponding to the beta-model is fractal with fractal dimension $d = 2/(\beta-1)$, confirming equation (15) of the main text.

Moreover, time dependence of the correlation function in the $N \to \infty$ limit is a function of a scaling dimensionless variable
\be
\tau = \frac{t}{\tilde{t}_N s^{\beta/\alpha}}.
\label{tau}
\ee
For $\tau \ll 1$ one expects the first terms of the series to be sufficient, and the resulting correlation function can be expressed in terms of a stretched exponent:
\be
G(s,t)_{\tau \ll 1} \sim \frac{a^2 s^{\beta-1}}{\Gamma(\alpha+1)} \left[
\exp \left(-\tau^{\alpha}\right)
+f(\beta)\Gamma(\alpha+1)-1\right]
\label{res_zero}
\ee
where
\be
f(\beta) = \sum_{k=1}^{\infty}(-1)^{k+1}\frac{\pi^{2k}}{(2k)!} \frac{1}{2k - \beta + 1}
\label{fbeta}
\ee

Now, for the behavior of \eq{result} at longer times, $\tau \gtrsim 1$ the fact that $N$ is not infinite becomes important. It is instructive to introduce $\epsilon = s/N$ and rewrite the integral \eq{f} in the form:
\be
I_{\alpha}(k, s, \tau) = \frac{1}{\beta} \left(\frac{1}{\tau^\alpha s^{\beta}}\right)^{g(k)}\left[\gamma_{\alpha}\left(g(k), \tau^\alpha \right)-\gamma_{\alpha}\left(g(k), \epsilon^{\beta} \tau^{\alpha}\right)\right]
\label{ifig}
\ee
where $g(k) = (2k+1) \beta^{-1} - 1$, and
\be
\gamma_{\alpha}(a, z) = \int_{0}^{z} x^{a-1} E_{\alpha}(-x) dx
\label{fig}
\ee
For $\alpha=1$ this function reduces to standard incomplete $\Gamma$-function (recall that $E_{1}(-x)=\exp(-x)$).

Consider the non-fractal case of $\alpha=1$ first. In this case it is known that for large $z$:
\be
\gamma(a, z) \approx \Gamma(a) - z^{a-1}\exp(-z)
\ee
For $\eps^{-\alpha/\beta} \gg \tau \gg 1$ the integral in \eq{nf} is controlled by the upper bound resulting in:
\be
I_{1}(k, s, \tau) \approx \frac{\Gamma(g(k))}{\beta}\left(\frac{1}{\tau s^{\beta}}\right)^{g(k)} -\frac{\exp\left(-\tau\right)}{\beta\tau}\frac{1}{s^{2k+1-\beta}}
\label{nf2}
\ee
In this intermediate regime, we may neglect the exponential term in \eq{nf2} and leave only $k=1$ term in the $k$-series, which leads the following expression for $G(s,\tau)$
\be
G(s,\tau)_{\frac{1}{\eps} \gg \tau \gg 1} \sim a^2 \Gamma\left(\frac{3 - \beta}{\beta}\right) s^{\beta-1} \tau^{-3 \beta^{-1} +1}.
\label{nf_int}
\ee

In turn, for $\tau \gg 1/\eps \gg 1$ $I_1$ is $s$-independent:
\be
I_{1}(k, s, \tau) \approx \frac{1}{\beta \tau \eps^{\beta}}N^{-\beta g(k)}\exp\left(-\eps^{\beta} \tau \right)
 = \frac{\tau_N}{\beta tN^{\beta(g(k)-1)}}\exp\left(-\frac{t}{\tau_N N^\beta}\right)
\label{nf3}
\ee
resulting in the following long-time asymptotic of the autocorrelation function:
\be
G(s,t)_{\tau \gg \eps^{-\beta}} \sim \frac{a^2 \tau_N}{t}
N^{2\beta-1}\exp\left(-\frac{t}{\tau_N N^\beta}\right)\left(1-\cos\frac{\pi s}{N}\right)
\ee

In order to proceed further in the case of $\alpha<1$, rewrite $\gamma_{\alpha}(a, z)$ in terms of the generalized Wright function\cite{Samko}:
\be
\gamma_{\alpha}(a, z) = {z^s} _2\psi_2\left[-z\bigg|
\begin{matrix}
(a, 1) & (1, 1) \\
(1, \alpha) & (1+a, 1)
\end{matrix} \right]
\ee
where, by definition:
\be
_p\psi_q\left[z\bigg|
\begin{matrix}
(a_1, \alpha_1) & (a_2, \alpha_2) & \cdots (a_p, \alpha_p) \\
(b_1, \beta_1) & (b_2, \beta_2) & \cdots (b_q, \beta_q)
\end{matrix} \right] = \sum_{k=0}^{\infty} \frac{\prod_{r=1}^{p} \Gamma(a_r + \alpha_r k)}{\prod_{r=1}^{q} \Gamma(b_r + \beta_r k)} \frac{z^k}{k!}
\ee
Applying the known results for the asymptotics of the generalized Wright function \cite{Paris2010}, one obtains the following approximation for $I_\alpha(k, s, \tau)$ at intermediate times $\eps^{-\beta/\alpha} \gg \tau \gg 1$:
\be
I_\alpha(k,s, \tau) \approx \dfrac{1}{\beta} s^{\beta-1-2k}
\left[\tau^{-\alpha g} A_1(g) + \tau^{-\alpha} A_2 (g) +2\tau ^{-1}\exp\left(-\left|\cos\dfrac{\pi}{\alpha}\right|\tau \right)
\cos\left(\dfrac{\pi}{\alpha} - \tau \sin\left(\dfrac{\pi}{\alpha}\right)
\right)\right]
\label{wright_asym}
\ee
where
\be
\begin{array}{rll}
A_1 (g)& =& \dfrac{\pi}{\sin\left(\pi g\right)}\dfrac{1}{\Gamma(1-\alpha g)} \medskip \\
A_2 (g) & = &\dfrac{1}{(g-1)\Gamma(1-\alpha)}
\end{array}
\ee
and we have omitted the $k$ dependence in $g(k)$ for brevity.

In the Newtonian case $\alpha=1$ the second prefactor $A_2(g)$ equals zero for all $k$, but for $\alpha<1$ both $A_1$ and $A_2$ are non-negligible and the comparative significance of respective terms in \eq{wright_asym} depends on $\beta$. As a result, there are two different asymptotic regimes depending on $\beta$: for $3 > \beta > 3/2$ (which is the most physically interesting case) the leading term is $\tau^{-\alpha g(1)}$, while for $3/2 > \beta > 1$ the dominant term is simply $\tau^{-\alpha}$. Note that latter values of $\beta$ correspond to very dense polymer conformations with fractal dimension $d>4$, and it is not very surprising that corresponding globules move around essentially as point-like particles \cite{Deng2009}.

As a result, at intermediate times and for $3 > \beta > 3/2$, the autocorrelation function reads:
\be
G(s,\tau)_{\eps^{-\beta/\alpha}\gg\tau\gg1} \sim a^2 s^{\beta-1} \frac{\Gamma(x)\Gamma(1-x)}{\Gamma(1-\alpha x)} \left(\frac{1}{\tau}\right)^{\alpha w}; \; \; w = g(1) = 3\beta^{-1} - 1.
\label{G_fract_interm}
\ee

In turn, for $3/2 > \beta > 1$:
\be
G(s,\tau)_{\eps^{-\beta/\alpha}\gg\tau\gg1}  \sim a^2 s^{\beta-1} \frac{f(2\beta)}{\Gamma(1-\alpha)} \frac{1}{\tau^\alpha}
\label{b_lt}
\ee
Finally, in the large time limit, $\tau \gg \eps^{-\beta/\alpha}$ there is a single regime for all $0 < \alpha < 1$:
\be
G(s,t)_{\tau \gg \eps^{-\beta/\alpha}} \sim \frac{a^2 N^{2\beta-1}}{\Gamma(1-\alpha)t^\alpha} q(\beta,\eps)
\label{b_llt}
\ee
where $q(\beta,x)$ is:
\be
q(\beta, x) = \sum_{k=1}^\infty \frac{(-1)^{k+1} (\pi x)^{2k}}{(2k)!} \frac{x^{2\beta-2k-1} - 1}{2k+1-2\beta} =
x^{2\beta-1}f(2\beta) - \sum_{k=1}^\infty \frac{(-1)^{k+1} (\pi x)^{2k}}{(2k)!} \frac{1}{2k+1-2\beta}
\ee

In $\eps \ll 1$ limit one may leave only $k=1$ terms in the latter sum and asymptotic of $q(\beta, x)$ at $x \to 0$
depends again on values of $\beta$:
\be
\begin{array}{rlll}
q(\beta, x)_{x \ll 1}& \approx & \frac{x^2}{2\beta-3}\; & \mbox{ for } 3/2 < \beta < 3 \medskip \\
q(\beta, x)_{x \ll 1} & \approx & x^{2\beta-1}f(2\beta)\; &  \mbox{ for } 1 < \beta < 3/2
\end{array}
\ee



\bibliography{joint_dynamics} 

\end{document}